\documentstyle[12pt]{article}
\begin{document}

\tolerance=5000

\def\cL{{\cal L}}
\def\be{\begin{equation}}
\def\ee{\end{equation}}
\def\bea{\begin{eqnarray}}
\def\eea{\end{eqnarray}}
\def\tr{{\rm tr}\, }
\def\nn{\nonumber \\}
\def\gd{g^\dagger}
\def\e{{\rm e}}
\newcommand{\inv}[1]{\left[#1\right]_{\mbox{inv}}}
\newcommand\DS{D \hskip -3mm / \ }
\def\ds{\left( 1 + {M \over \lambda}
\e^{\lambda(\sigma^- - \sigma^+ )}
\right)}

\  \hfill
\begin{minipage}{2.5cm}
NDA-FP-36 \\
July 1997 \\
\end{minipage}

\

\vfill

\begin{center}

{\large\bf 
TRACE ANOMALY INDUCED EFFECTIVE ACTION AND HAWKING
RADIATION FOR 2d DILATONIC SUPERGRAVITY}

\vfill

{\large\sc Shin'ichi NOJIRI}\footnote{
e-mail : nojiri@cc.nda.ac.jp}
and
{\large\sc Sergei D. ODINTSOV$^{\spadesuit}$}\footnote{
e-mail :
odintsov@quantum.univalle.edu.co, \\
odintsov@kakuri2-pc.phys.sci.hiroshima-u.ac.jp}
\vfill

{\large\sl Department of Mathematics and Physics \\
National Defence Academy \\
Hashirimizu Yokosuka 239, JAPAN}

{\large\sl $\spadesuit$
Tomsk Pedagogical University \\
634041 Tomsk, RUSSIA \\
and \\
Dep.de Fisica \\
Universidad del Valle \\
AA25360, Cali, COLOMBIA \\
}

\vfill

{\bf ABSTRACT}

\end{center}

We construct the theory of 2d dilatonic supergravity(SG)
with matter and dilaton supermultiplets coupled to
dilaton functions. Trace anomaly and induced effective action
for matter supermultiplet are calculated (what gives also
large-$N$ effective action for dilatonic SG).
Study of black holes and Hawking radiation which turns
out to be zero in supersymmetric CGHS model with dilaton
coupled matter is presented. In the same way one can
study spherically symmetric collapse for other 4d SG
using simplified 2d approach.

\ 


\newpage
There are different motivations to study 2d black holes
\cite{Rev}.
First of all, such models are string inspired ones. Second,
using spherically symmetric reduction anzats of 4d Einstein-scalar
 theory one is left with the action of 2d dilatonic gravity
with dilaton coupled matter \cite{8}. 
Hence, 4d spherical collapse
maybe understood in terms of 2d dilatonic gravity. There
were recently attempts to study trace anomaly and induced
effective action for 2d dilaton coupled scalars 
\cite{6,4,5}
(what was motivated by appearence of dilaton coupled scalars
after reduction) as well as related research of 2d black holes
and Hawking radiation with account of quantum effects of
dilaton coupled scalars.

It would be of interest to present supersymmetric generalization
of such investigation as spherically symmetric reduction of
4d SG with matter leads to 2d dilatonic SG with 
dilaton coupled
matter supermultiplet. That will be the purpose 
of present work-
to find anomaly induced effective action in 2d 
dilatonic SG with matter.

At first we are going to construct the action
of 2d dilatonic supergravity with dilaton
supermultiplet and
with matter supermultiplet.
In order to construct the Lagrangian of two-dimensional dilatonic
supergravity, we use the component formulation
 of ref.\cite{HUY} (for introduction, 
see book\cite{GSRG}).

In this paper, all the scalar fields are real and
all the spinor fields are Majorana spinors.

We introduce dilaton multiplet $\Phi=(\phi,\chi,F)$
and matter multiplet $\Sigma_i=(a_i,\chi_i,G_i)$,
which has the conformal weight $\lambda=0$, and
the curvature multiplet $W$

\be
\label{W}
W=\left(S,\eta,-S^2-{1 \over 2}R-{1 \over 2}
\bar\psi^\mu\gamma^\nu\psi_{\mu\nu}
+{1 \over 4}\bar\psi^\mu\psi_\mu\right)\ .
\ee
Here $R$ is the scalar curvature, for other notations 
see \cite{HUY}.

Then the general action of 2d dilatonic supergravity
is given in terms of general functions of the
dilaton $C(\phi)$, $Z(\phi)$, $f(\phi)$ and
$V(\phi)$
 as follows
\bea
\label{lag}
&&\cL=-\inv{C(\Phi)\otimes W} + \inv{V(\Phi)} \nn
&& +{1 \over 2}
\inv{\Phi\otimes\Phi\otimes T_P(Z(\Phi))}
-\inv{Z(\Phi)\otimes\Phi \otimes
T_P(\Phi)} \nn
&& +\sum_{i=1}^N
\left\{{1 \over 2}
\inv{\Sigma_i\otimes\Sigma_i\otimes T_P(f(\Phi))}
-\inv{f(\Phi)\otimes \Sigma_i\otimes T_P(\Sigma_i)}
\right\}\ .
\eea
$T_P(Z)$ is called the kinetic multiplet and
$\inv{Z}$ expresses the invariant Lagrangian.
Hence, we constructed the classical action for 2d
dilatonic supergravity with dilaton and matter supermultiplets.

We now study the trace anomaly
and effective action in large-$N$ approximation
for the 2d dilatonic supergravity.
We consider only bosonic background below
as it will be sufficient
for our purposes (study of black hole type solutions).

On the bosonic
background where the dilatino $\chi$ and
the Rarita-Schwinger fields vanish, one can
show that
the gravity and dilaton part of the Lagrangian have
the following form:
\bea
\label{gdlag}
&&\inv{C(\Phi)\otimes W}
= e\left[-C(\phi)\left(S^2 +{1 \over 2}R\right)
-C'(\phi)FS\right] \ ,\nn
&& \inv{\Phi\otimes\Phi\otimes T_P(Z(\Phi))}
=e\left[\phi^2\tilde\Box(Z(\phi))
+ 2Z'(\phi)\phi F^2\right] \ ,\nn
&& \inv{Z(\Phi)\otimes\Phi \otimes T_P(\Phi)}
=e\left[Z(\phi)\phi\tilde\Box\phi
+ Z'(\phi)\phi F^2 + Z(\phi)F^2 \right] \ ,\nn
&& \inv{V(\Phi)} = e\left[ V'(\phi)F  + SV(\phi)\right]
\ .
\eea
For the matter part we obtain
\bea
\label{bg}
&&\inv{f(\Phi)\otimes \Sigma_i\otimes T_P(\Sigma_i)} 
=e\left[ f(\phi)(a_i\tilde\Box a_i
-\bar\xi_i\tilde\DS\xi_i ) + f'(\phi)Fa_iG_i
+ f(\phi)G_i^2\right] \nn
&& \inv{\Sigma_i\otimes\Sigma_i\otimes T_P(f(\Phi))} 
=e\left[a_i^2\tilde\Box(f(\phi))
+ 2f'(\phi)Fa_iG_i\right]\ .
\eea

Using the equations of motion with respect to
the auxilliary fields $S$, $F$, $G_i$,
on the bosonic background one can show that
\be
\label{bgaf}
S={C'(\phi)V'(\phi) - 2V(\phi)Z(\phi)
\over {C'}^2(\phi) + 4C(\phi) Z (\phi)} , \ 
F={C'(\phi) V(\phi)+ 2 C(\phi)V'(\phi)
\over {C'}^2(\phi) + 4C(\phi) Z (\phi)} , \ 
G_i=0.
\ee
We will be interested later in the supersymmetric extension \cite{NO} of
the CGHS model \cite{12} as in specifical example for study
of black holes and Hawking radiation. For such a model
\be
\label{CGHS}
C(\phi)=2\e^{-2\phi}\ ,\ \ Z(\phi)=4\e^{-2\phi}\ ,
\ \ V(\phi)=4\e^{-2\phi}\ ,
\ee
we find
\be
\label{CGHSaf}
S=0\ ,\ \ F=-\lambda\ ,\ \ G_i=0\ .
\ee

Using (\ref{bg}), we can show that on bosonic background
\bea
\label{lag2}
&& \sum_{i=1}^N
\left\{{1 \over 2}
\inv{\Sigma_i\otimes\Sigma_i\otimes T_P(f(\Phi))}
-\inv{f(\Phi)\otimes \Sigma_i\otimes T_P(\Sigma_i)}\right\} \\
&&=ef(\phi)\sum_{i=1}^N
(g^{\mu\nu}\partial_\mu a_i \partial_\nu a_i
+\bar\xi_i\gamma^\mu\partial_\mu\xi_i
-f(\phi)G_i^2) 
+ \left(
\begin{array}{c}\mbox{\small total divergence} \\
\mbox{\small terms} \end{array}\right) \nonumber
\eea
Here we have used the fact that
$\bar\xi_i \gamma_5 \xi=0$
for the Majorana spinor $\xi_i$.

Let us start now the investigation of effective action
in above theory. It is clearly seen that theory
(\ref{lag2}) is conformally invaiant on the
gravitational background under discussion.
Then using standard methods, we can prove that theory
with matter multiplet $\Sigma_i$ is superconformally
invariant theory.
First of all, one can find the trace anomaly $T$
for the theory (\ref{lag2}) on gravitational
background using the following relation
\be
\label{gamma}
\Gamma_{div}={1 \over n-2}\int d^2x \sqrt g b_2\ , \
\ \ \ T=b_2
\ee
where $b_2$ is $b_2$ coefficient of
Schwinger-De Witt expansion and $\Gamma_{div}$
is one-loop effective action.
The calculation of $\Gamma_{div}$ (\ref{gamma}) for
the quantum theory with Lagrangian (\ref{lag2}) has
been done some time ago in ref.\cite{6}.
Using results of this work, we find
\bea
\label{tracean}
T&=&{1 \over 24\pi}\left\{{3 \over 2}NR
- 3N\left({f'' \over f} - {{f'}^2 \over 2f^2}\right)
(\nabla^\lambda\phi)(\nabla_\lambda\phi)
- 3N{f' \over f}\Delta\phi\right\}\ .
\eea
It is remarkable that Majorana spinors do not give
the contribution to the dilaton dependent terms in
trace anomaly as it was shown in \cite{6}.
They only alter the coefficient of curvature
term in $T$ (\ref{tracean}).
Hence, except the coefficient of curvature term in
$T$ (\ref{tracean}), the trace anomaly (\ref{tracean})
coincides with the correspondent expression for dilaton
coupled scalar \cite{4}.
Note also that for particular case
$f(\phi)=\e^{-2\phi}$ the trace anomaly for
dilaton coupled scalar has been recently calculated in
refs.\cite{5}.

Making now the conformal transformation of the
metric $g_{\mu\nu}\rightarrow \e^{2\sigma}g_{\mu\nu}$
in the trace anomaly, and using relation:
\be
\label{TW}
T={1 \over \sqrt g}{\delta \over \delta \sigma}
W[\sigma]
\ee
one can find anomaly induced action $W[\sigma]$.
In the covariant, non-local form it may be found as
following:
\be
\label{qc}
W=-{1 \over 2}\int d^2x \sqrt{g} \left[
{N \over 32\pi}R{1 \over \Delta}R
-{N \over 16\pi}{{f'}^2 \over f^2}
\nabla^\lambda \phi
\nabla_\lambda \phi {1 \over \Delta}R
-{N \over 8\pi}\ln f R \right]\ .
\ee
Hence, we got the anomaly induced effective action
for dilaton coupled matter multiplet in the
external dilaton-gravitational background.
We should note that the same action $W$ (\ref{qc})
gives the one-loop large-$N$ effective action in the
quantum theory of supergravity with matter (\ref{lag})
(i.e., when all fields are quantized).

We can now rewrite $W$ in a supersymmetric way.
In order to write down the effective action expressing
the trace anomaly, we need the supersymmetric extention
of ${ 1 \over \Delta}R$. The extension is given by
using the inverse kinetic multiplet in \cite{MNSN},
or equivalently by introducing two auxiliary field
$\Theta=(t,\theta,T)$ and $\Upsilon=(u,\upsilon,U)$.
We can now construct the following action
\be
\label{TU}
\inv{\Theta\otimes (T_P(\Upsilon)-W)}\ .
\ee
The $\Theta$-equation of motion tells that, in the
superconformal gauge
\be
\label{scgauge}
e^a_\mu =\e^\rho \delta^a_\mu\ (e=\e^\rho)\ ,
\ \ \psi_\mu=\gamma_\mu \psi\
(\bar\psi_\mu=-\bar\psi\gamma_\mu)
\ee
we find
\be
\label{rp}
u\sim \rho \sim -{1 \over 2\Delta}R
\ , \ \ \ \upsilon\sim \psi\ .
\ee
Then we obtain
\bea
\label{actions}
&& \sqrt g R{1 \over \Delta}R\sim
4\inv{W\otimes \Upsilon} \nn
&& \sim -\inv{
\Sigma_i\otimes\Sigma_i\otimes T_P\left(
\left({{f'}^2(\Phi) \over f(\Phi)} - f''(\Phi)
\right)\right)\Upsilon} \nn
&& \hskip 1cm +2\inv{
\left({{f'}^2(\Phi) \over f(\Phi)} - f''(\Phi)\right)
\otimes\Upsilon\otimes\Sigma_i
\otimes T_P(\Sigma_i)} \nn
&& \sqrt g {{f'}^2(\phi) \over f^2(\phi)}
\nabla_\lambda\phi \nabla^\lambda \phi
{1 \over \Delta}R \nn
&& \sim -\inv{\Phi\otimes\Phi
\otimes T_P\left({{f'}^2(\Phi) \over f^2(\Phi)}
\otimes \Upsilon \right)}
+ 2\inv{{{f'}^2(\Phi) \over f^2(\Phi)}
\otimes \Upsilon \otimes \Phi \otimes T_P(\Phi)} \nn
&& \sqrt g \ln f(\phi) \, R\sim
-2\inv{\ln f(\Phi)\otimes W}\ .
\eea
That finishes the construction of large-$N$ effective
action for 2d dilatonic supergravity with matter in 
supersymmetric form.

We now discuss the particular 2d
dilatonic supergravity model which represents
the supersymmetric extension of CGHS model. Note
that as a matter we use dilaton coupled matter supermultiplet.
We would like to estimate back-reaction of such matter
supermultiplet to black holes and Hawking radiation
working in large-$N$ approximation.
Since we are interesting in the vacuum (black hole)
solution, we consider the background where
matter fields, the Rarita-Schwinger field and
dilatino vanish.

In the superconformal gauge
the equations of motion can be obtained by the variation
over $g^{\pm\pm}$, $g^{\pm\mp}$ and $\phi$
\bea
\label{eqnpp}
&&0=T_{\mu\nu}^c+ T_{\mu\nu}^q,\hskip 1cm
0={\cal P}^c + {\cal P}^q \nn
&&T_{\pm\pm}^c=\e^{-2\phi}\left(4\partial_\pm \rho
\partial_\pm\phi - 2 \left(\partial_\pm\phi\right)^2
\right) \nn
&&T_{\pm\pm}^q = {N \over 8}\left( \partial_\pm^2 \rho
- \partial_\pm\rho \partial_\pm\rho \right) \nn
&& +{N \over 8} \left\{
\left(
\partial_\pm h(\phi) \partial_\pm h(\phi) \right)
\rho+{1 \over 2}{\partial_\pm \over \partial_\mp}
\left( \partial_\pm h(\phi)
\partial_\mp h(\phi) \right)\right\} \nn
&& +{N \over 8}\left\{
-2 \partial_\pm \rho \partial_\pm h(\phi)
+\partial_\pm^2 h(\phi) \right\} 
 + {N \over 64}{\partial_\pm \over \partial_\mp}
\left( h'(\phi)^2F^2 \right) + t^\pm(x^\pm) \nn
&& T_{\pm\mp}^c
=\e^{-2\phi}\left(2\partial_+
\partial_- \phi -4 \partial_+\phi\partial_-\phi
- \lambda^2 \e^{2\rho}\right) \nn
&&T_{\pm\mp}^q = -{N \over 8}\partial_+\partial_- \rho
-{N \over 16}\partial_+ h(\phi)
\partial_- \tilde \phi
-{N \over 8}\partial_+\partial_-h(\phi) \nn
&& -{N \over 64}h'(\phi)F^2
+\left({N \over 16}US
+ {N \over 2}(-h(\phi)S^2 h'(\phi)FS\right)\e^{2\rho}
\nn
&&{\cal P}^c= \e^{-2\phi}\left(-4\partial_+
\partial_- \phi +4 \partial_+\phi\partial_-\phi
+2\partial_+ \partial_- \rho
+ \lambda^2 \e^{2\rho}\right) \nn
&&{\cal P}^q =-{Nf' \over f}\left\{
{1 \over 16}\partial_+(\rho \partial_-h(\phi))
+{1 \over 16}\partial_-(\rho \partial_+h(\phi))
-{1 \over 8}\partial_+\partial_-\rho \right\} \ .
\eea
Here $h(\phi)\equiv\ln f(\phi)$ and 
$t^\pm(x^\pm)$ is a function which is determined
by the boundary condition.
Note that there is, in general, a contribution from
the auxilliary fields to $T_{\pm\mp}$ besides the
contribution from the trace anomaly.

In large-$N$ limit, where
classical part can be ignored, the field equations
become simpler
\be
\label{eqnpp2}
0=T_{\mu\nu}^q\ ,\hskip 1cm 0={\cal P}^q\ .
\ee
Here we used the $\Theta$-equation and the equations
for the auxilliary fields $S$ and $F$.
The function $t^\pm(x^\pm)$ in (\ref{eqnpp2})
can be absorbed into
the choice of the coordinate and we can choose
$t^\pm(x^\pm)=0$.
We can show that the general solutions of
(\ref{eqnpp2}) are given by
\be
\label{phi}
h(\phi)= \int d\rho {1 \pm
\sqrt{1 + \rho} \over \rho} \ ,\hskip 1cm
\rho=-1 +
\left(\rho^+(x^+) + \rho^-(x^-)\right)^{2 \over 3}\ .
\ee
Here $\rho^\pm$ is an arbitrary function of
$x^\pm = t \pm x$.
The scalar curvature is given by
\be
\label{sR}
R=8\e^{-2\rho}\partial_+\partial_-\rho \nn
= -{4\e^{-2\left\{
-1+\left(\rho^+(x^+)+\rho^-(x^-)\right)^{2 \over 3}
\right\}} {\rho^+}'(x^+){\rho^-}'(x^-) \over
\left(\rho^+(x^+)+\rho^-(x^-)
\right)^{{4 \over 3}}}\ .
\ee
Note that when $\rho^+(x^+)+\rho^-(x^-)=0$,
there is a curvature singularity.
Especially if we choose
\be
\label{Krus}
\rho^+(x^+)={r_0 \over x^+}\ ,\ \
\rho^-(x^-)=-{x^- \over r_0}
\ee
there are curvature singularities at $x^+x^-=r_0^2$ and
horizon at $x^+=0$ or $x^-=0$.
Hence we got black hole solution in the model
under discussion.
The asymptotic flat
regions are given by $x^+\rightarrow +\infty$ ($x^-<0$)
or $x^-\rightarrow -\infty$ ($x^+>0$).

We now consider the Hawking radiation
in the bosonic background where
the fermionic fields vanish.
We investigate the case that
\be
\label{BHch}
f(\phi)=\e^{\alpha\phi}\ \  (h(\phi)=\alpha\phi)\ .
\ee
Substituting the classical black hole solution
which appeared in the original CGHS model \cite{12}
\be
\label{sws}
\rho=-{1 \over 2}\ln \left(1 + {M \over \lambda}
\e^{\lambda (\sigma^--\sigma^+ )} \right)\ ,\hskip 1cm
\phi=- {1 \over 2}\ln \left( {M \over \lambda}
+ \e^{\lambda(\sigma^+ - \sigma^-)} \right)
\ee
(Here $M$ is the mass of the black hole and we used asymptotic flat coordinates.)
into the quantum part of the energy momentum tensor
$T_{\mu\nu}^q$ in (\ref{eqnpp2})
and using eq.(\ref{CGHSaf}),
we find the explicit expressions for quantum
energy-momentum tensor.

Then when $\sigma^+\rightarrow +\infty$, the energy
momentum tensor behaves as
\be
\label{asT}
T^q_{+-}\rightarrow 0
 \ ,\ \ 
T^q_{\pm\pm}\rightarrow {N\lambda^2 \over 16}
\alpha^2 + t^\pm(\sigma^\pm)\ .
\ee
In order to evaluate $t^\pm(\sigma^\pm)$,
we impose a boundary condition that
there is no incoming energy.
This condition requires that $T^q_{++}$
should vanish at the past null infinity
($\sigma^-\rightarrow +\infty$) and if we assume $t^-(\sigma^-)$ is
black hole mass independent,
$T^q_{--}$ should also vanish at
the past horizon ($\sigma^+\rightarrow -\infty$) after taking
$M \rightarrow 0$ limit.
Then we find
\be
\label{t-}
t^\pm(\sigma^\pm)=-{N\lambda^2\alpha^2 \over 16}
\ee
and one obtains
\be
\label{rad}
T^q_{--}\rightarrow 0
\ee
at the future null infinity
($\sigma^+\rightarrow +\infty$).
Eqs.(\ref{asT}) and (\ref{rad}) might
tell that there is no
the Hawking radiation in the dilatonic supergravity model
under discussion when quantum back-reaction of
matter supermultiplet in large-$N$ approach
is taken into account.
This might be the result of the positive energy theorem \cite{PS}.
If the Hawking radiation is the positive and mass independent, the energy
of the system becomes unbounded below. It could be also that since
we work in strong coupling regime new methods to study Hawking
radiation should be developed.

We would like to thank R. Bousso, I.L. Buchbinder, 
S.W. Hawking, S.J. Gates and K. Stelle
for helpful remarks.

\end{document}